\documentclass[footinbib, aps,pre,twocolumn,showpacs,amsfonts,amssymb,floatfix, superscriptaddress, reprint]{revtex4-1}
\usepackage[T1]{fontenc}
\usepackage[utf8]{inputenc}
\usepackage{float}
\usepackage{graphicx}
\usepackage{amssymb}
\usepackage{amsmath}
\usepackage{enumerate}

\makeatletter

\newcommand*{\sinc}{\mathop{\mathrm{sinc}}}

\date{May 15, 2017} 

\makeatother

\begin{document}

\title{Giant Kovacs-Like Memory Effect for Active Particles}

\author{Rüdiger Kürsten}
\affiliation{Institut für Physik, Universität Greifswald, Felix-Hausdorff-Str. 6, 17489 Greifswald}
\author{Vladimir Sushkov}
\affiliation{Hochschule für angewandte Wissenschaften München, Fakultät für angewandte Naturwissenschaften und Mechatronik, Lothstr. 34, 80335 München}
\author{Thomas Ihle}
\affiliation{Institut für Physik, Universität Greifswald, Felix-Hausdorff-Str. 6, 17489 Greifswald}

\begin{abstract}
Dynamical properties of a Vicsek-like gas of self-propelled particles are investigated by means of kinetic theory and agent based simulations.
While memory effects have been observed in disordered systems,
we show that they also occur in active matter systems.
In particular, we find that the system exhibits a giant Kovacs-like memory effect that is much larger than predicted by a generic linear theory.
Based on a separation of time scales we develop a nonlinear theory to explain this effect.
We apply this theory to driven granular gases and propose further applications to spin glasses.
\end{abstract}
\maketitle

Collective behavior is a crucial aspect in the rapidly growing field of active matter
\cite{EWG15, MJRLPRS13, VZ12, Ramaswamy10}.
Active particles under consideration can be animals like e.g. insects, fish or birds \cite{CKFL05}, interacting robots \cite{RCN14}, or microscopic objects like e.g. bacteria \cite{GRLC14}, nano-dimers \cite{TK10} or Janus particles \cite{JYS10}.
There is shared belief, that, on a macroscopic level, active particle systems can be described by a minimal set of hydrodynamic fields \cite{peshkov_12b}.
There has been large emphasis in deriving field equations using different approaches \cite{bertin_06, bertin_09, peshkov_12a, weber_13, peshkov_12b, peruani_08, gross_13, gross_16, ihle_11, chou_12, ihle_15, ihle_16, baskaran_08a, baskaran_08b, CK14}.
For polar particles with ferromagnetic alignment interactions the description by Toner and Tu's seminal equations \cite{TT95,TT98} is well established.
Steady states of homogeneous solutions have also been studied in detail \cite{RLI14, LSD15, KI17}.
However, memory effects and dynamical properties of active matter far from its steady state are largely unexplored.

To study non-stationary properties and possible history-dependencies we consider the following prototype situation:
Imagine a substance that, at time $t=0$, is in equilibrium with a heat bath of temperature $T_1$, and assume that the temperature of the heat bath could be changed instantaneously.
Suppose, we want to heat the substance to a higher temperature $T_{f}>T_1$.
This could be achieved by simply adjusting the heat bath to the desired final temperature $T_{f}$. 
After a certain time, the substance will have relaxed to equilibrium at $T_f$.
Trying to speed up this process one could initially set the heat bath to a higher temperature $T_2>T_f$ and switch the temperature of the heat bath to $T_f$ after a particular waiting time $t_w$.
Because the amount of heat transfered to the system until time $t=t_w$ is increased by this protocol, we intuitively expect the system to reach the desired temperature faster.
This procedure is related to a measuring protocol introduced by Kovacs et. al \cite{Kovacs63, KAHR79}.
In addition to the aforementioned steps they chose the waiting time $t_w$ in a particular way.
Considering an observable $\Psi$, the waiting time was chosen such that at the moment of temperature switching $t_w$, the observable $\Psi$ has the same value that it has in equilibrium at temperature $T_f$.
Thus $\Psi(t_w)=\Psi(t \rightarrow \infty)$.
In the case of \cite{Kovacs63, KAHR79}, $\Psi$ is the volume and here it is the polar order parameter. 

If the order parameter $\Psi(t)$ enslaved all other degrees of freedom as implicitly assumed in Toner-Tu theories \cite{TT95, TT98, TTU98, TTR05, peshkov_12a, Toner12}, the system would be in equilibrium already at $t=t_w$ and remain unchanged for the rest of the experiment.
If, however, at $t=t_w$ there are additional hidden degrees of freedom that are not perfectly enslaved, they will couple to the observable $\Psi(t)$ such that  
it first departs from and later relaxes back towards its equilibrium value.
This is known as the Kovacs-effect.
The state of the hidden degrees of freedom depends on the entire history of the process.
Thus the presence of a significant Kovacs-hump demonstrates that memory effects are crucial.
In recent years the Kovacs temperature protocol was applied e.g. on glass systems \cite{BB02, CLL04, AS04, RP14}, molecular liquids \cite{MS04}, but also on non-equilibrium systems like driven granular gases \cite{PT14, BGMB14, TP14}.
In the latter case the role of the temperature is replaced by a driving force.
In this Letter, we investigate the Kovacs-effect for a system of interacting polar active particles that are subject to angular noise, where the noise strength $\eta$ takes over the role of the temperature.

Surprisingly, we observe a giant Kovacs-hump that reveals the presence of strong memory effects.
A reduced description via density and polar order parameter in the manner of Toner and Tu is not incorporating such history-dependent effects and thus is not sufficient to describe this particular
polar active gas. 
We find that the Kovacs protocol does not lead to a speedup of the relaxation towards the final steady state.
On the contrary, due to the giant Kovacs-effect a slower relaxation is observed.
Speaking in analogy to the thermal system described above, this means that the temporal increase of the heat bath temperature to $T_2$ results in a slower heating of the system.
This result is highly counterintuitive and contradicts a linear theory on the Kovacs-effect \cite{PB10}.

In this Letter, 
we develop a quantitative nonlinear theory relating the giant Kovacs-effect to two types of relaxation curves by employing a much more general framework than the present active matter system.
With this central result we achieve an intuitive understanding of the effect.
It becomes clear that a separation of time scales in one of the relaxation curves is responsible for the appearance of the giant Kovacs-hump.

Here, we consider a two-dimensional Vicsek-like model for self-propelled particles with bounded confidence interactions, introduced in \cite{RLI14}.
Bounded confidence interactions have been studied first in the context of opinion formation models \cite{DNAW00, HK02} to mimic the tendency to ignore others with opposite opinions.
One motivation for such interactions are experiments on the collective motion of \textit{Bacillus subtilis} \cite{lu_13}, where the authors propose weaker interactions between cells with increasing orientation difference.
In the Vicsek model, all particles move at constant speed in individual directions. 
For a streaming period of unit time they evolve ballistically and afterwards they interact instantaneously.
Each particle adopts the mean direction of motion of all particles that are no further away than some interaction distance and that move in a direction that differs by no more than the angle $\alpha$ from the particles own direction.
Note, that in that way each particle interacts at least with itself.
Then all particle directions are disturbed by a random deviation $\xi$ that is drawn for each particle independently from the interval $[-\eta/2, \eta/2]$.

Considering only spatially homogeneous solutions one obtains a time evolution equation for the angular distribution of the particle directions.
Assuming molecular chaos and a low particle density $M$ this equation is explicitly given in angular Fourier space \cite{KI17, RLI14} by a hierarchy of equations
\begin{align}
	x_{k}(t+1)=& \lambda_k\Big\{ A_k x_k + \sum_{q=1}^{\infty}x_q \big[ B_k x_{|k-q|} + C_{k} x_{k+q}   \big]   \Big\}
	\label{eq:timeevolutionfourier}
\end{align}
with $x_0=\frac{1}{2\pi}$ at all times. 
The coefficients $\lambda_k(\eta, M)$, $A_k(M, \alpha)$, $B_{k}(M, \alpha)$ and $C_{k}(M, \alpha)$ can be found in the supplemental material \footnote{\label{fn:supp}See supplemental material at pages 6ff.}.
In real space, the angular distribution is then given by $p(\theta) = \sum_{k=0}^{\infty} x_k \cos(k\theta)$.
For practical computations we have to truncate the Fourier series after finitely many terms,
setting
$x_k=0$ for $k>n$.
Here, we use $n=100$ or $n=200$. We denote the vector of Fourier modes $(x_1, \dots, x_n)$ by $\mathbf{x}$.
As an observable we consider the polar order parameter $\Psi:= \langle \cos(\theta)\rangle = \pi x_1$.
If the system is perfectly ordered, all particles move in the same direction and $\Psi=1$.
If, in contrast, the directions of all particles are completely random, we have $\Psi=0$.

\begin{figure}
	\includegraphics[width=0.47\textwidth]{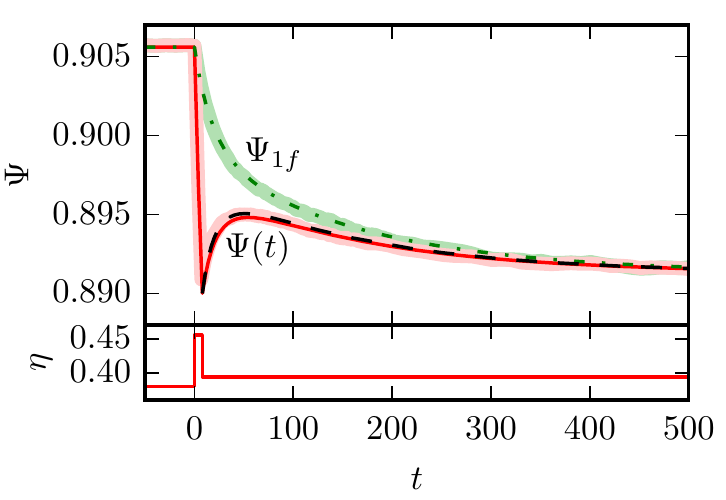}
	\caption{The relaxation curve $\Psi_{1f}(t)$ (dash-dotted line) and the Kovacs-hump $\Psi(t)$ (solid red line) obtained from the kinetic theory \eqref{eq:timeevolutionfourier} are in good agreement with agent-based simulations (light green and light red thick line, respectively). 
	The linear theory \eqref{eq:lintheory} (dashed black line) describes the Kovacs-effect well for short $t_w=8$.
	System parameters are $\alpha=0.47\pi$, $M=0.2$, $\eta_1=0.3797$, $\eta_2=0.4553$, $\eta_f=0.3940$ \cite{Note2}. The lower part shows the noise strength according to the Kovacs-protocoll.
	\label{fig:simul1}}
\end{figure}

Linearizing in the change of temperature, which corresponds to a change in $\eta$ in our case, the following relation between the Kovacs-hump $\Psi(t)$ and the relaxation curve ${\Psi}_{1f}$ was derived in Ref. \cite{PB10} by means of a Master-equation approach
\begin{align}
	{\Psi}(t)&= \frac{1}{1-\gamma}{\Psi}_{1f}(t) - \frac{\gamma}{1-\gamma} {\Psi}_{1f}(t-t_{w}),
	\label{eq:lintheory}
	\\
	\gamma &= \frac{{\Psi}_{1f}(t_{w})-{\Psi}^{*}_{\eta_f}}{{\Psi}_{1f}(0)-{\Psi}^{*}_{\eta_f}}.
	\label{eq:gamma}
\end{align}
By $\Psi_{1f}(t)$ we denote the order parameter, when the system is prepared at $t=0$ in the steady state of noise strength $\eta_1$ and for $t>0$ the noise strength is switched to $\eta_f$. 
The steady state value of the order parameter at noise strength $\eta$ is denoted by $\Psi^{*}_{\eta}$.
As long as the relaxation curve ${\Psi}_{1f}$ is monotone, it follows from Eq.~\eqref{eq:gamma} that $\gamma\in (0,1)$, and hence
\begin{align}
	{\Psi}(t) \begin{cases} 
		> {\Psi}_{1f} \text{ if } {\Psi}_{1f} \text{ is increasing} \\
		< {\Psi}_{1f} \text{ if } {\Psi}_{1f} \text{ is decreasing}. 
	\end{cases}
	\label{eq:conditionlintheory}
\end{align}
We evaluated the time evolution map \eqref{eq:timeevolutionfourier} numerically and changed the noise strength according to the Kovacs-protocol.
The results are displayed in Fig.~\ref{fig:simul1}.
In the same plot we also present results of agent based simulations, where positions $z_j$ and directions $\theta_j$ of the $j$th particle are updated according to the streaming rule
\begin{align}
	z_{j}(t+1) = z_{j}(t) + v_0 \binom{\sin(\theta_j)}{\cos(\theta_j)}
	\label{eq:streaming}
\end{align}
with particle speed $v_0$, and the collision rule
\begin{align}
	\tilde{\theta}_j &= \Phi_j + \xi_j, 
	\qquad
	\Phi_j = \arg \Bigg[ \sum_{l\in \{j\} } \exp(i \theta_l)   \Bigg],
	\label{eq:collision2}
\end{align}
where $\tilde{\theta}_j$ denotes the post-collisional coordinate of the $j$th particle and $\xi_j$ are independent random variables that are drawn uniformly from the interval $[-\eta/2, \eta/2]$.
The set $\{ j \}$ contains the indices of all particles that interact with the $j$th particle according to the bounded confidence rule.
There is good agreement between agent based simulations and kinetic theory \eqref{eq:timeevolutionfourier}.
We clearly find that the Kovacs-effect is present in this system.
In Fig.~\ref{fig:simul1} where the waiting time $t_w=8$ is very small, the linear theory \eqref{eq:lintheory} agrees well with the data.
\phantom{
\footnote{\label{foot:1}Since the density $M=0.2$ is not that small the description by the kinetic theory \eqref{eq:timeevolutionfourier} is not perfect and the steady state of the real system is slightly different.\label{1} To compare kinetic theory and simulations we used slightly different noise strengths for the simulation, such that the steady state value of the order parameter coincides with the kinetic theory for all three noise strengths. Simulation parameters are $\eta_1=0.38$, $\eta_2=0.46$, $\eta_f=0.395$.}
}

\begin{figure}
	\includegraphics[width=0.47\textwidth]{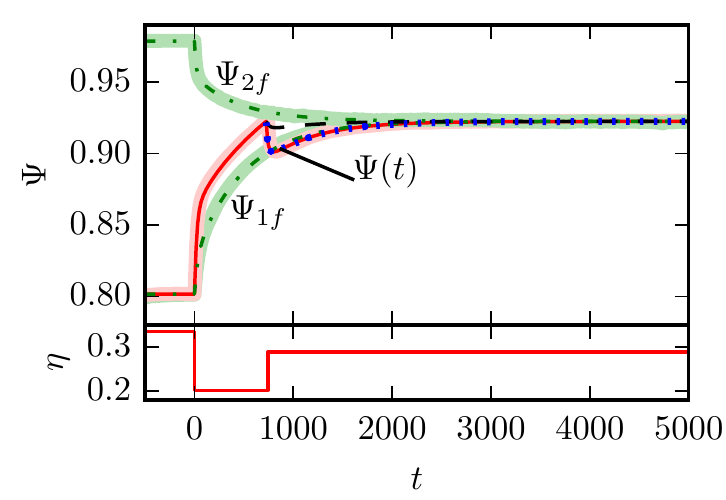}
	\caption{The relaxation curves $\Psi_{1f}(t)$, $\Psi_{2f}(t)$ (dash-dotted line) and the Kovacs-hump $\Psi(t)$ (solid red line) obtained from the kinetic theory \eqref{eq:timeevolutionfourier} are in good agreement with agent-based simulations (light green and light red thick line, respectively). 
	The linear theory \eqref{eq:lintheory} (dashed black line) fails to describe the Kovacs-hump whereas the nonlinear theory \eqref{eq:kovacshump} with time shift $\hat{t}=332$ (dotted blue line) agrees well.
Parameters are $\alpha=0.35\pi$, $M=0.2$, $t_{w}=742$, $\eta_1=0.3354$, $\eta_2=0.2017$, $\eta_f=0.2884$ \cite{Note3}.
	\label{fig:giant}}
\end{figure}

However, for a different parameter set, in particular a smaller $\alpha=0.35 \pi$, and for the inverse Kovacs-protocol with $\eta_1>\eta_f>\eta_2$, displayed in Fig.\ref{fig:giant}, we immediately recognize that the Kovacs-hump $\Psi(t)$ and the relaxation curve $\Psi_{1f}(t)$ intersect.
That means the relaxation towards the final steady state under the Kovacs-protocol is slower than the direct relaxation from $\eta_1$ to $\eta_f$.
This surprising result clearly violates condition \eqref{eq:conditionlintheory} which is a consequence of the linear theory \eqref{eq:lintheory}.
The Kovacs-hump is giant compared to the predictions of the linear theory (dashed black line in Fig.~\ref{fig:giant}).
In the supplemental material \footnotemark[1] we rederive Eq.~\eqref{eq:lintheory} for time discrete dynamical systems linearizing the time evolution map as a function of the system state and of the noise strength. We argue that this linearized theory is only applicable if the waiting time is short.
Thus, in the present case, the linear theory is not sufficient to describe the system and nonlinear effects are crucial.

\phantom{
\footnote{Analogously to \cite{Note2} simulation noise strengths differ minimally: $\eta_1=0.34$, $\eta_2=0.197$, $\eta_f=0.29$.}
}

Developing a nonlinear theory, we find that the vector of Fourier modes $\mathbf{x}(t)$ obtained from the Kovacs-protocol for $t\ge t_{w}$ is related to both relaxation curves $\mathbf{x}_{1f}$ and $\mathbf{x}_{2f}$ by the following central result 
\begin{align}
	\mathbf{x}(t)=\mathbf{x}_{2f}(t-t_w) + \mathbf{x}_{1f}(t-\hat{t})-\mathbf{x}_{2f}(t-\hat{t})
	\label{eq:kovacshump}
\end{align}
We derive this equation not by linearizing the time evolution Eq.~\eqref{eq:timeevolutionfourier} itself but only the change of the system's state after $t_w$.
Furthermore, we do not linearize in the change of the noise strength but we take a strong nonlinear dependence of the relaxation speed on the noise strength into account, see supplemental material \footnotemark[1] for details.
The first vector component of Eq.~\eqref{eq:kovacshump} yields the corresponding relation for the order parameter $\Psi=\pi x_1$ but Eq.~\eqref{eq:kovacshump} is more general, yielding a relation for all Fourier modes.
The time shift $\hat{t}$ in Eq.~\eqref{eq:kovacshump} depends on the change of the relaxation speed when $\eta$ is switched (see \footnotemark[1] for more details).
The right-hand-side of Eq.~\eqref{eq:kovacshump} is displayed as the blue dotted line in Fig.~\ref{fig:giant}.
We see that it coincides very well with the Kovacs-hump. We discuss a plot of the second Fourier mode of the same process in \footnotemark[1].

With the help of Eq.~\eqref{eq:kovacshump} we can understand why the Kovacs-effect is so large.
We find that the derivative of $\Psi(t)$ at $t=t_w$ is given by $\Psi'_{2f}(0) + \Psi'_{1f}(t_w-\hat{t}) - \Psi'_{2f}(t_w-\hat{t})$.
In the present case, the relaxation curve $\Psi_{2f}$ is decaying very fast in the beginning and both relaxation curves are decaying much slower at later times, cf. Fig.~\ref{fig:giant}.
Since $|\Psi'_{1f}(t)|$ and $|\Psi'_{2f}(t)|$ are of comparable size, the term $\Psi'_{2f}(0)$ is dominant.
That means a fast relaxation of $\Psi_{2f}$ at $t=0$ leads to a fast and therefore strong change of $\Psi(t)$ in a short period after $t=t_w$.
Thus, whenever $|\Psi'_{1f}|\sim |\Psi'_{2f}|$ and $\Psi_{2f}$ is relaxing very fast in the beginning and much slower at later times we expect a giant Kovacs-effect.

The derivation of Eq.~\eqref{eq:kovacshump} is not system specific but it is valid for a wide class of systems with the following two properties.
(i) In the relaxation dynamics there must be a separation of time scales. In particular there has to be one mode that relaxes much slower than all others such that at $t_w$ we can assume that all modes but one are already completely relaxed. (ii) Furthermore, we need to assume that this slow mode is not too sensitive to changes in $\eta$ and $t$ such that the slowest relaxation mode for $\eta_2$ is approximately equal to the one at $\eta_f$.

For the present system these properties are numerically verified (see supplemental material \footnotemark[1] for more details).
They can also be understood intuitively.
Assume a large population of particles moves into direction $\theta=0$.
Then they interact only with others that move in a direction from the interval $[-\alpha, \alpha]$.
For $\alpha=0.35 \pi$ this interval is a little larger than $2\pi/3$.
Particles that have directions outside this interval can not interact with the first population.
Therefore, it is possible that there is a second, relatively large, population of particles that move into the opposite direction $\theta=\pi$.
Then, the interaction intervals of both populations are disjoint and all particles can interact only with either the first or the second population.
Particles can be driven from one population to the other only by noise.
For small noise strength this process is very slow, in particular because the interaction of particles within the same population acts against this noise driven mechanism.
In contrast, the concentration of particles that belong to one population due to aligning interactions is a much faster process.

\begin{figure}
	\includegraphics{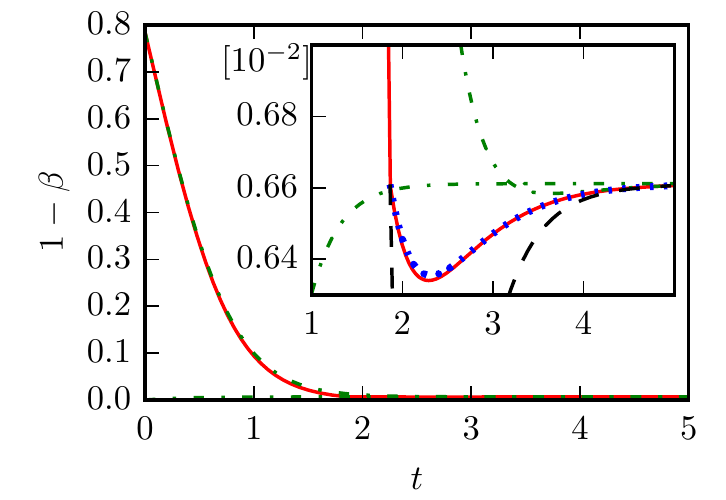}
	\caption{Anomalous Kovacs-effect for a driven granular gas.
	The order parameter $\beta$ is related to the granular temperature.
	Displayed are relaxation curves (green dash-dotted lines) and the Kovacs-hump (red solid line).
	The Kovacs-effect is very small and shown in the inset in units of $10^{-2}$.
	The nonlinear theory (blue dotted line), Eq.~\eqref{eq:kovacshump} with $\hat{t}$=0, describes the anomalous effect well, the linear theory (black dashed line), Eq.~\eqref{eq:lintheory}, fails.\label{fig:granular}}
\end{figure}

Thus, we identified a very slow relaxation process in the system: the noise driven reorientation of a group of particles moving in the opposite direction than the majority.
This relaxation process is very robust against moderate changes in noise strength. 
Although the relaxation speed depends strongly on the noise strength, the mechanism remains the slowest dynamics as long as the noise is not too strong.
In the derivation of Eq.~\eqref{eq:kovacshump} we neither use that the initial configuration is a steady state nor that the order parameter $\Psi(t_w)= \Psi^{}_{}(t\rightarrow \infty)$ (see \footnotemark[1]).
Thus Eq.~\eqref{eq:kovacshump} is applicable to more general protocols than the Kovacs experiment.
We conjecture that Eq.~\eqref{eq:kovacshump} might also be valid for completely different types of systems with slow dynamics like e.g. driven granular gases or spin glasses.
This conjecture can be verified theoretically or experimentally by measuring the Kovacs-hump as well as the relaxation curves $\Psi_{1f}$ and $\Psi_{2f}$ in different types of systems. The time shift $\hat{t}$ might be obtained from theoretical calculations specific to the system or simply from a fit of experimental or simulation data. 
For one specific granular gas we verify Eq.~\eqref{eq:kovacshump} explicitly.
In Refs.~\cite{PT14, TP14} the dynamics of a driven granular gas is given by ordinary differential equations, Eqs.~(6a) and (6b) in Ref. \cite{PT14}.
Evaluating these equations numerically we obtain the Kovacs-hump $\Psi(t)$ as well as the relaxation curves $\Psi_{1f}(t)$ and $\Psi_{2f}(t)$, such that we can test the validity of Eq.~\eqref{eq:kovacshump}.
In order to satisfy conditions (i) and (ii) we must choose intermediate and final noise intensities close to each other.
In that way we assure that the waiting time $t_w$ is long and that the slowest relaxation mode can be approximated as constant.
In Fig.~\ref{fig:granular} we show the anomalous Kovacs-effect for the driven granular gas of Refs.\cite{PT14, TP14}. We find that also the anomalous effect is described well by Eq.~\eqref{eq:kovacshump} whereas the linear theory \eqref{eq:lintheory} fails completely. 
We can understand why the Kovacs-hump has the opposite sign by investigating the derivative of Eq.~\eqref{eq:kovacshump}.
In the present case $|\Psi'_{1f}| \gg |\Psi'_{2f}|$ and hence the term $\Psi'_{1f}(t-\hat{t})$ is dominant, causing the anomalous effect.

For spin glasses, property (i) is not realistic but it can be replaced by an alternative assumption.
Consider the coherence length $l$ of a spin glass such that length scales smaller than $l$ are equilibrated and length scales larger than $l$ are frozen.
The coherence length depends on the age of the system. 
Turning it around, one finds that the age of the system typically grows faster than exponentially with the coherence length \cite{BDHV01, BB02}.
Translating this picture into the framework of relaxing eigenmodes we assume the following property: 
(i*) After the waiting time $t_w$ some eigenmodes have relaxed completely, some others have not relaxed at all and there is a single mode which is partely relaxed.
If (i*) holds instead of (i), the derivation of Eq.~\eqref{eq:kovacshump} remains valid.
The validity of Eq.~\eqref{eq:kovacshump} for spin glasses could be verified in the future.
However, equilibrium states might be difficult to prepare both in experiments and in simulations.
Fortunately, as argued above, Eq.~\eqref{eq:kovacshump} also holds when the initial state is not an equilibrium state.
Unfortunately the measurement of the relaxation curve $\mathbf{x}_{2f}$ requires the preparation of the equilibrium state belonging to $\eta=\eta_2$.
However, Eq.~\eqref{eq:kovacshump} yields still insights on the system even if $\mathbf{x}_{2f}$ can not be measured.

The time shift $\hat{t}$ in Eq.~\eqref{eq:kovacshump} does not depend on the initial configuration but only on the intermediate and final noise strength and the waiting time $t_w$. Hence rewriting Eq.~\eqref{eq:kovacshump} as
\begin{align}
	\mathbf{x}(t) -\mathbf{x}_{1f}(t-\hat{t})=\mathbf{x}_{2f}(t-t_w)-\mathbf{x}_{2f}(t-\hat{t})
	\label{eq:kovacshumprewrite}
\end{align}
we find that the right-hand-side is independent on the initial configuration.
Therefore also the left-hand-side must be identical for different initial configurations.
Even if equilibrium states can not be prepared in an experimental or simulation setup, the quantities on the left-hand-side of Eq.~\eqref{eq:kovacshumprewrite} can still be measured.
Comparing them for different initial configurations Eq.~\eqref{eq:kovacshumprewrite} can be verified and one could extract information on the relaxation from one equilibrium state to another without preparing any equilibrium state at all.

In summary, we observe a giant Kovacs-memory-effect in an active matter system, both, in agent based simulations and in kinetic theory.
The effect is unexpected as it contradicts a well-known linear theory.
Furthermore, it shows that the Turner-Tu equations are not sufficient to describe the dynamics of the polar active gas investigated here.
We develop a quantitative nonlinear theory that connects the Kovacs-hump with two relaxation curves.
We conclude that the giant effect is caused by a fast relaxation on short time scales and a slow relaxation on large time scales.
We also apply the nonlinear theory to a driven granular gas where it succeeds to quantitatively describe the anomalous Kovacs-effect.
We further propose experimental validations and applications to spin glasses and other disordered systems.

We thank A. Prados for detailed insights in driven granular gases. 
We thank T. Voigtmann for valuable discussions.

\bibliography{literatur.bib}

\begin{thebibliography}{51}%
\makeatletter
\providecommand \@ifxundefined [1]{%
 \@ifx{#1\undefined}
}%
\providecommand \@ifnum [1]{%
 \ifnum #1\expandafter \@firstoftwo
 \else \expandafter \@secondoftwo
 \fi
}%
\providecommand \@ifx [1]{%
 \ifx #1\expandafter \@firstoftwo
 \else \expandafter \@secondoftwo
 \fi
}%
\providecommand \natexlab [1]{#1}%
\providecommand \enquote  [1]{``#1''}%
\providecommand \bibnamefont  [1]{#1}%
\providecommand \bibfnamefont [1]{#1}%
\providecommand \citenamefont [1]{#1}%
\providecommand \href@noop [0]{\@secondoftwo}%
\providecommand \href [0]{\begingroup \@sanitize@url \@href}%
\providecommand \@href[1]{\@@startlink{#1}\@@href}%
\providecommand \@@href[1]{\endgroup#1\@@endlink}%
\providecommand \@sanitize@url [0]{\catcode `\\12\catcode `\$12\catcode
  `\&12\catcode `\#12\catcode `\^12\catcode `\_12\catcode `\%12\relax}%
\providecommand \@@startlink[1]{}%
\providecommand \@@endlink[0]{}%
\providecommand \url  [0]{\begingroup\@sanitize@url \@url }%
\providecommand \@url [1]{\endgroup\@href {#1}{\urlprefix }}%
\providecommand \urlprefix  [0]{URL }%
\providecommand \Eprint [0]{\href }%
\providecommand \doibase [0]{http://dx.doi.org/}%
\providecommand \selectlanguage [0]{\@gobble}%
\providecommand \bibinfo  [0]{\@secondoftwo}%
\providecommand \bibfield  [0]{\@secondoftwo}%
\providecommand \translation [1]{[#1]}%
\providecommand \BibitemOpen [0]{}%
\providecommand \bibitemStop [0]{}%
\providecommand \bibitemNoStop [0]{.\EOS\space}%
\providecommand \EOS [0]{\spacefactor3000\relax}%
\providecommand \BibitemShut  [1]{\csname bibitem#1\endcsname}%
\let\auto@bib@innerbib\@empty
\bibitem [{\citenamefont {Elgeti}\ \emph {et~al.}(2015)\citenamefont {Elgeti},
  \citenamefont {Winkler},\ and\ \citenamefont {Gompper}}]{EWG15}%
  \BibitemOpen
  \bibfield  {author} {\bibinfo {author} {\bibfnamefont {J.}~\bibnamefont
  {Elgeti}}, \bibinfo {author} {\bibfnamefont {R.~G.}\ \bibnamefont {Winkler}},
  \ and\ \bibinfo {author} {\bibfnamefont {G.}~\bibnamefont {Gompper}},\
  }\href@noop {} {\bibfield  {journal} {\bibinfo  {journal} {Rep. Prog. Phys.}\
  }\textbf {\bibinfo {volume} {78}},\ \bibinfo {pages} {056601} (\bibinfo
  {year} {2015})}\BibitemShut {NoStop}%
\bibitem [{\citenamefont {Marchetti}\ \emph {et~al.}(2013)\citenamefont
  {Marchetti}, \citenamefont {Joanny}, \citenamefont {Ramaswamy}, \citenamefont
  {Liverpool}, \citenamefont {Prost}, \citenamefont {Rao},\ and\ \citenamefont
  {Simha}}]{MJRLPRS13}%
  \BibitemOpen
  \bibfield  {author} {\bibinfo {author} {\bibfnamefont {M.}~\bibnamefont
  {Marchetti}}, \bibinfo {author} {\bibfnamefont {J.}~\bibnamefont {Joanny}},
  \bibinfo {author} {\bibfnamefont {S.}~\bibnamefont {Ramaswamy}}, \bibinfo
  {author} {\bibfnamefont {T.}~\bibnamefont {Liverpool}}, \bibinfo {author}
  {\bibfnamefont {J.}~\bibnamefont {Prost}}, \bibinfo {author} {\bibfnamefont
  {M.}~\bibnamefont {Rao}}, \ and\ \bibinfo {author} {\bibfnamefont {R.~A.}\
  \bibnamefont {Simha}},\ }\href@noop {} {\bibfield  {journal} {\bibinfo
  {journal} {Rev. Mod. Phys.}\ }\textbf {\bibinfo {volume} {85}},\ \bibinfo
  {pages} {1143} (\bibinfo {year} {2013})}\BibitemShut {NoStop}%
\bibitem [{\citenamefont {Vicsek}\ and\ \citenamefont {Zafeiris}(2012)}]{VZ12}%
  \BibitemOpen
  \bibfield  {author} {\bibinfo {author} {\bibfnamefont {T.}~\bibnamefont
  {Vicsek}}\ and\ \bibinfo {author} {\bibfnamefont {A.}~\bibnamefont
  {Zafeiris}},\ }\href@noop {} {\bibfield  {journal} {\bibinfo  {journal}
  {Phys. Rep.}\ }\textbf {\bibinfo {volume} {517}},\ \bibinfo {pages} {71}
  (\bibinfo {year} {2012})}\BibitemShut {NoStop}%
\bibitem [{\citenamefont {Ramaswamy}(2010)}]{Ramaswamy10}%
  \BibitemOpen
  \bibfield  {author} {\bibinfo {author} {\bibfnamefont {S.}~\bibnamefont
  {Ramaswamy}},\ }\href@noop {} {\bibfield  {journal} {\bibinfo  {journal}
  {Annu. Rev. Cond. Matt. Phys.}\ }\textbf {\bibinfo {volume} {1}},\ \bibinfo
  {pages} {323} (\bibinfo {year} {2010})}\BibitemShut {NoStop}%
\bibitem [{\citenamefont {Couzin}\ \emph {et~al.}(2005)\citenamefont {Couzin},
  \citenamefont {Krause}, \citenamefont {Franks},\ and\ \citenamefont
  {Levin}}]{CKFL05}%
  \BibitemOpen
  \bibfield  {author} {\bibinfo {author} {\bibfnamefont {I.~D.}\ \bibnamefont
  {Couzin}}, \bibinfo {author} {\bibfnamefont {J.}~\bibnamefont {Krause}},
  \bibinfo {author} {\bibfnamefont {N.~R.}\ \bibnamefont {Franks}}, \ and\
  \bibinfo {author} {\bibfnamefont {S.~A.}\ \bibnamefont {Levin}},\ }\href@noop
  {} {\bibfield  {journal} {\bibinfo  {journal} {Nature}\ }\textbf {\bibinfo
  {volume} {433}},\ \bibinfo {pages} {513} (\bibinfo {year}
  {2005})}\BibitemShut {NoStop}%
\bibitem [{\citenamefont {Rubenstein}\ \emph {et~al.}(2014)\citenamefont
  {Rubenstein}, \citenamefont {Cornejo},\ and\ \citenamefont {Nagpal}}]{RCN14}%
  \BibitemOpen
  \bibfield  {author} {\bibinfo {author} {\bibfnamefont {M.}~\bibnamefont
  {Rubenstein}}, \bibinfo {author} {\bibfnamefont {A.}~\bibnamefont {Cornejo}},
  \ and\ \bibinfo {author} {\bibfnamefont {R.}~\bibnamefont {Nagpal}},\
  }\href@noop {} {\bibfield  {journal} {\bibinfo  {journal} {Science}\ }\textbf
  {\bibinfo {volume} {345}},\ \bibinfo {pages} {795} (\bibinfo {year}
  {2014})}\BibitemShut {NoStop}%
\bibitem [{\citenamefont {Gachelin}\ \emph {et~al.}(2014)\citenamefont
  {Gachelin}, \citenamefont {Rousselet}, \citenamefont {Lindner},\ and\
  \citenamefont {Clement}}]{GRLC14}%
  \BibitemOpen
  \bibfield  {author} {\bibinfo {author} {\bibfnamefont {J.}~\bibnamefont
  {Gachelin}}, \bibinfo {author} {\bibfnamefont {A.}~\bibnamefont {Rousselet}},
  \bibinfo {author} {\bibfnamefont {A.}~\bibnamefont {Lindner}}, \ and\
  \bibinfo {author} {\bibfnamefont {E.}~\bibnamefont {Clement}},\ }\href@noop
  {} {\bibfield  {journal} {\bibinfo  {journal} {New J. Phys.}\ }\textbf
  {\bibinfo {volume} {16}},\ \bibinfo {pages} {025003} (\bibinfo {year}
  {2014})}\BibitemShut {NoStop}%
\bibitem [{\citenamefont {Tao}\ and\ \citenamefont {Kapral}(2010)}]{TK10}%
  \BibitemOpen
  \bibfield  {author} {\bibinfo {author} {\bibfnamefont {Y.-G.}\ \bibnamefont
  {Tao}}\ and\ \bibinfo {author} {\bibfnamefont {R.}~\bibnamefont {Kapral}},\
  }\href@noop {} {\bibfield  {journal} {\bibinfo  {journal} {Soft Matter}\
  }\textbf {\bibinfo {volume} {6}},\ \bibinfo {pages} {756} (\bibinfo {year}
  {2010})}\BibitemShut {NoStop}%
\bibitem [{\citenamefont {Jiang}\ \emph {et~al.}(2010)\citenamefont {Jiang},
  \citenamefont {Yoshinaga},\ and\ \citenamefont {Sano}}]{JYS10}%
  \BibitemOpen
  \bibfield  {author} {\bibinfo {author} {\bibfnamefont {H.-R.}\ \bibnamefont
  {Jiang}}, \bibinfo {author} {\bibfnamefont {N.}~\bibnamefont {Yoshinaga}}, \
  and\ \bibinfo {author} {\bibfnamefont {M.}~\bibnamefont {Sano}},\ }\href
  {\doibase 10.1103/PhysRevLett.105.268302} {\bibfield  {journal} {\bibinfo
  {journal} {Phys. Rev. Lett.}\ }\textbf {\bibinfo {volume} {105}},\ \bibinfo
  {pages} {268302} (\bibinfo {year} {2010})}\BibitemShut {NoStop}%
\bibitem [{\citenamefont {Peshkov}\ \emph
  {et~al.}(2012{\natexlab{a}})\citenamefont {Peshkov}, \citenamefont {Aranson},
  \citenamefont {Bertin}, \citenamefont {Chat{\'e}},\ and\ \citenamefont
  {Ginelli}}]{peshkov_12b}%
  \BibitemOpen
  \bibfield  {author} {\bibinfo {author} {\bibfnamefont {A.}~\bibnamefont
  {Peshkov}}, \bibinfo {author} {\bibfnamefont {I.~S.}\ \bibnamefont
  {Aranson}}, \bibinfo {author} {\bibfnamefont {E.}~\bibnamefont {Bertin}},
  \bibinfo {author} {\bibfnamefont {H.}~\bibnamefont {Chat{\'e}}}, \ and\
  \bibinfo {author} {\bibfnamefont {F.}~\bibnamefont {Ginelli}},\ }\href@noop
  {} {\bibfield  {journal} {\bibinfo  {journal} {Phys. Rev. Lett.}\ }\textbf
  {\bibinfo {volume} {109}},\ \bibinfo {pages} {268701} (\bibinfo {year}
  {2012}{\natexlab{a}})}\BibitemShut {NoStop}%
\bibitem [{\citenamefont {Bertin}\ \emph {et~al.}(2006)\citenamefont {Bertin},
  \citenamefont {Droz},\ and\ \citenamefont {Gr{\'e}goire}}]{bertin_06}%
  \BibitemOpen
  \bibfield  {author} {\bibinfo {author} {\bibfnamefont {E.}~\bibnamefont
  {Bertin}}, \bibinfo {author} {\bibfnamefont {M.}~\bibnamefont {Droz}}, \ and\
  \bibinfo {author} {\bibfnamefont {G.}~\bibnamefont {Gr{\'e}goire}},\
  }\href@noop {} {\bibfield  {journal} {\bibinfo  {journal} {Phys. Rev. E}\
  }\textbf {\bibinfo {volume} {74}},\ \bibinfo {pages} {022101} (\bibinfo
  {year} {2006})}\BibitemShut {NoStop}%
\bibitem [{\citenamefont {Bertin}\ \emph {et~al.}(2009)\citenamefont {Bertin},
  \citenamefont {Droz},\ and\ \citenamefont {Gr{\'e}goire}}]{bertin_09}%
  \BibitemOpen
  \bibfield  {author} {\bibinfo {author} {\bibfnamefont {E.}~\bibnamefont
  {Bertin}}, \bibinfo {author} {\bibfnamefont {M.}~\bibnamefont {Droz}}, \ and\
  \bibinfo {author} {\bibfnamefont {G.}~\bibnamefont {Gr{\'e}goire}},\
  }\href@noop {} {\bibfield  {journal} {\bibinfo  {journal} {J. Phys. A}\
  }\textbf {\bibinfo {volume} {42}},\ \bibinfo {pages} {445001} (\bibinfo
  {year} {2009})}\BibitemShut {NoStop}%
\bibitem [{\citenamefont {Peshkov}\ \emph
  {et~al.}(2012{\natexlab{b}})\citenamefont {Peshkov}, \citenamefont {Ngo},
  \citenamefont {Bertin}, \citenamefont {Chat{\'e}},\ and\ \citenamefont
  {Ginelli}}]{peshkov_12a}%
  \BibitemOpen
  \bibfield  {author} {\bibinfo {author} {\bibfnamefont {A.}~\bibnamefont
  {Peshkov}}, \bibinfo {author} {\bibfnamefont {S.}~\bibnamefont {Ngo}},
  \bibinfo {author} {\bibfnamefont {E.}~\bibnamefont {Bertin}}, \bibinfo
  {author} {\bibfnamefont {H.}~\bibnamefont {Chat{\'e}}}, \ and\ \bibinfo
  {author} {\bibfnamefont {F.}~\bibnamefont {Ginelli}},\ }\href@noop {}
  {\bibfield  {journal} {\bibinfo  {journal} {Phys. Rev. Lett.}\ }\textbf
  {\bibinfo {volume} {109}},\ \bibinfo {pages} {098101} (\bibinfo {year}
  {2012}{\natexlab{b}})}\BibitemShut {NoStop}%
\bibitem [{\citenamefont {Weber}\ \emph {et~al.}(2013)\citenamefont {Weber},
  \citenamefont {Th{\"u}roff},\ and\ \citenamefont {Frey}}]{weber_13}%
  \BibitemOpen
  \bibfield  {author} {\bibinfo {author} {\bibfnamefont {C.~A.}\ \bibnamefont
  {Weber}}, \bibinfo {author} {\bibfnamefont {F.}~\bibnamefont {Th{\"u}roff}},
  \ and\ \bibinfo {author} {\bibfnamefont {E.}~\bibnamefont {Frey}},\
  }\href@noop {} {\bibfield  {journal} {\bibinfo  {journal} {New J. Phys.}\
  }\textbf {\bibinfo {volume} {15}},\ \bibinfo {pages} {045014} (\bibinfo
  {year} {2013})}\BibitemShut {NoStop}%
\bibitem [{\citenamefont {Peruani}\ \emph {et~al.}(2008)\citenamefont
  {Peruani}, \citenamefont {Deutsch},\ and\ \citenamefont
  {B{\"a}r}}]{peruani_08}%
  \BibitemOpen
  \bibfield  {author} {\bibinfo {author} {\bibfnamefont {F.}~\bibnamefont
  {Peruani}}, \bibinfo {author} {\bibfnamefont {A.}~\bibnamefont {Deutsch}}, \
  and\ \bibinfo {author} {\bibfnamefont {M.}~\bibnamefont {B{\"a}r}},\
  }\href@noop {} {\bibfield  {journal} {\bibinfo  {journal} {Eur. Phys. J.
  Spec. Top.}\ }\textbf {\bibinfo {volume} {157}},\ \bibinfo {pages} {111}
  (\bibinfo {year} {2008})}\BibitemShut {NoStop}%
\bibitem [{\citenamefont {Gro{\ss}mann}\ \emph {et~al.}(2013)\citenamefont
  {Gro{\ss}mann}, \citenamefont {Schimansky-Geier},\ and\ \citenamefont
  {Romanczuk}}]{gross_13}%
  \BibitemOpen
  \bibfield  {author} {\bibinfo {author} {\bibfnamefont {R.}~\bibnamefont
  {Gro{\ss}mann}}, \bibinfo {author} {\bibfnamefont {L.}~\bibnamefont
  {Schimansky-Geier}}, \ and\ \bibinfo {author} {\bibfnamefont
  {P.}~\bibnamefont {Romanczuk}},\ }\href@noop {} {\bibfield  {journal}
  {\bibinfo  {journal} {New J. Phys.}\ }\textbf {\bibinfo {volume} {15}},\
  \bibinfo {pages} {085014} (\bibinfo {year} {2013})}\BibitemShut {NoStop}%
\bibitem [{\citenamefont {Gro{\ss}mann}\ \emph {et~al.}(2016)\citenamefont
  {Gro{\ss}mann}, \citenamefont {Peruani},\ and\ \citenamefont
  {B{\"a}r}}]{gross_16}%
  \BibitemOpen
  \bibfield  {author} {\bibinfo {author} {\bibfnamefont {R.}~\bibnamefont
  {Gro{\ss}mann}}, \bibinfo {author} {\bibfnamefont {F.}~\bibnamefont
  {Peruani}}, \ and\ \bibinfo {author} {\bibfnamefont {M.}~\bibnamefont
  {B{\"a}r}},\ }\href@noop {} {\bibfield  {journal} {\bibinfo  {journal} {Phys.
  Rev. E}\ }\textbf {\bibinfo {volume} {94}},\ \bibinfo {pages} {050602}
  (\bibinfo {year} {2016})}\BibitemShut {NoStop}%
\bibitem [{\citenamefont {Ihle}(2011)}]{ihle_11}%
  \BibitemOpen
  \bibfield  {author} {\bibinfo {author} {\bibfnamefont {T.}~\bibnamefont
  {Ihle}},\ }\href@noop {} {\bibfield  {journal} {\bibinfo  {journal} {Phys.
  Rev. E}\ }\textbf {\bibinfo {volume} {83}},\ \bibinfo {pages} {030901}
  (\bibinfo {year} {2011})}\BibitemShut {NoStop}%
\bibitem [{\citenamefont {Chou}\ \emph {et~al.}(2012)\citenamefont {Chou},
  \citenamefont {Wolfe},\ and\ \citenamefont {Ihle}}]{chou_12}%
  \BibitemOpen
  \bibfield  {author} {\bibinfo {author} {\bibfnamefont {Y.-L.}\ \bibnamefont
  {Chou}}, \bibinfo {author} {\bibfnamefont {R.}~\bibnamefont {Wolfe}}, \ and\
  \bibinfo {author} {\bibfnamefont {T.}~\bibnamefont {Ihle}},\ }\href@noop {}
  {\bibfield  {journal} {\bibinfo  {journal} {Phys. Rev. E}\ }\textbf {\bibinfo
  {volume} {86}},\ \bibinfo {pages} {021120} (\bibinfo {year}
  {2012})}\BibitemShut {NoStop}%
\bibitem [{\citenamefont {Ihle}(2015)}]{ihle_15}%
  \BibitemOpen
  \bibfield  {author} {\bibinfo {author} {\bibfnamefont {T.}~\bibnamefont
  {Ihle}},\ }\href@noop {} {\bibfield  {journal} {\bibinfo  {journal} {Eur.
  Phys. J. Spec. Top.}\ }\textbf {\bibinfo {volume} {224}},\ \bibinfo {pages}
  {1303} (\bibinfo {year} {2015})}\BibitemShut {NoStop}%
\bibitem [{\citenamefont {Ihle}(2016)}]{ihle_16}%
  \BibitemOpen
  \bibfield  {author} {\bibinfo {author} {\bibfnamefont {T.}~\bibnamefont
  {Ihle}},\ }\href@noop {} {\bibfield  {journal} {\bibinfo  {journal} {J. Stat.
  Mech.}\ }\textbf {\bibinfo {volume} {2016}},\ \bibinfo {pages} {083205}
  (\bibinfo {year} {2016})}\BibitemShut {NoStop}%
\bibitem [{\citenamefont {Baskaran}\ and\ \citenamefont
  {Marchetti}(2008{\natexlab{a}})}]{baskaran_08a}%
  \BibitemOpen
  \bibfield  {author} {\bibinfo {author} {\bibfnamefont {A.}~\bibnamefont
  {Baskaran}}\ and\ \bibinfo {author} {\bibfnamefont {M.~C.}\ \bibnamefont
  {Marchetti}},\ }\href@noop {} {\bibfield  {journal} {\bibinfo  {journal}
  {Phys. Rev. Lett.}\ }\textbf {\bibinfo {volume} {101}},\ \bibinfo {pages}
  {268101} (\bibinfo {year} {2008}{\natexlab{a}})}\BibitemShut {NoStop}%
\bibitem [{\citenamefont {Baskaran}\ and\ \citenamefont
  {Marchetti}(2008{\natexlab{b}})}]{baskaran_08b}%
  \BibitemOpen
  \bibfield  {author} {\bibinfo {author} {\bibfnamefont {A.}~\bibnamefont
  {Baskaran}}\ and\ \bibinfo {author} {\bibfnamefont {M.~C.}\ \bibnamefont
  {Marchetti}},\ }\href@noop {} {\bibfield  {journal} {\bibinfo  {journal}
  {Phys. Rev. E}\ }\textbf {\bibinfo {volume} {77}},\ \bibinfo {pages} {011920}
  (\bibinfo {year} {2008}{\natexlab{b}})}\BibitemShut {NoStop}%
\bibitem [{\citenamefont {Chepizhko}\ and\ \citenamefont
  {Kulinskii}(2014)}]{CK14}%
  \BibitemOpen
  \bibfield  {author} {\bibinfo {author} {\bibfnamefont {O.}~\bibnamefont
  {Chepizhko}}\ and\ \bibinfo {author} {\bibfnamefont {V.}~\bibnamefont
  {Kulinskii}},\ }\href@noop {} {\bibfield  {journal} {\bibinfo  {journal}
  {Physica A}\ }\textbf {\bibinfo {volume} {415}},\ \bibinfo {pages} {493}
  (\bibinfo {year} {2014})}\BibitemShut {NoStop}%
\bibitem [{\citenamefont {Toner}\ and\ \citenamefont {Tu}(1995)}]{TT95}%
  \BibitemOpen
  \bibfield  {author} {\bibinfo {author} {\bibfnamefont {J.}~\bibnamefont
  {Toner}}\ and\ \bibinfo {author} {\bibfnamefont {Y.}~\bibnamefont {Tu}},\
  }\href@noop {} {\bibfield  {journal} {\bibinfo  {journal} {Phys. Rev. Lett.}\
  }\textbf {\bibinfo {volume} {75}},\ \bibinfo {pages} {4326} (\bibinfo {year}
  {1995})}\BibitemShut {NoStop}%
\bibitem [{\citenamefont {Toner}\ and\ \citenamefont {Tu}(1998)}]{TT98}%
  \BibitemOpen
  \bibfield  {author} {\bibinfo {author} {\bibfnamefont {J.}~\bibnamefont
  {Toner}}\ and\ \bibinfo {author} {\bibfnamefont {Y.}~\bibnamefont {Tu}},\
  }\href@noop {} {\bibfield  {journal} {\bibinfo  {journal} {Phys. Rev. E}\
  }\textbf {\bibinfo {volume} {58}},\ \bibinfo {pages} {4828} (\bibinfo {year}
  {1998})}\BibitemShut {NoStop}%
\bibitem [{\citenamefont {Romensky}\ \emph {et~al.}(2014)\citenamefont
  {Romensky}, \citenamefont {Lobaskin},\ and\ \citenamefont {Ihle}}]{RLI14}%
  \BibitemOpen
  \bibfield  {author} {\bibinfo {author} {\bibfnamefont {M.}~\bibnamefont
  {Romensky}}, \bibinfo {author} {\bibfnamefont {V.}~\bibnamefont {Lobaskin}},
  \ and\ \bibinfo {author} {\bibfnamefont {T.}~\bibnamefont {Ihle}},\
  }\href@noop {} {\bibfield  {journal} {\bibinfo  {journal} {Phys. Rev. E}\
  }\textbf {\bibinfo {volume} {90}},\ \bibinfo {pages} {063315} (\bibinfo
  {year} {2014})}\BibitemShut {NoStop}%
\bibitem [{\citenamefont {Lam}\ \emph {et~al.}(2015)\citenamefont {Lam},
  \citenamefont {Schindler},\ and\ \citenamefont {Dauchot}}]{LSD15}%
  \BibitemOpen
  \bibfield  {author} {\bibinfo {author} {\bibfnamefont {K.-D. N.~T.}\
  \bibnamefont {Lam}}, \bibinfo {author} {\bibfnamefont {M.}~\bibnamefont
  {Schindler}}, \ and\ \bibinfo {author} {\bibfnamefont {O.}~\bibnamefont
  {Dauchot}},\ }\href@noop {} {\bibfield  {journal} {\bibinfo  {journal} {J.
  Stat. Mech.}\ }\textbf {\bibinfo {volume} {2015}},\ \bibinfo {pages} {P10017}
  (\bibinfo {year} {2015})}\BibitemShut {NoStop}%
\bibitem [{\citenamefont {K{\"u}rsten}\ and\ \citenamefont
  {Ihle}(2017)}]{KI17}%
  \BibitemOpen
  \bibfield  {author} {\bibinfo {author} {\bibfnamefont {R.}~\bibnamefont
  {K{\"u}rsten}}\ and\ \bibinfo {author} {\bibfnamefont {T.}~\bibnamefont
  {Ihle}},\ }\href@noop {} {\bibfield  {journal} {\bibinfo  {journal} {J. Stat.
  Mech.}\ }\textbf {\bibinfo {volume} {2017}},\ \bibinfo {pages} {033202}
  (\bibinfo {year} {2017})}\BibitemShut {NoStop}%
\bibitem [{\citenamefont {Kovacs}(1963)}]{Kovacs63}%
  \BibitemOpen
  \bibfield  {author} {\bibinfo {author} {\bibfnamefont {A.~J.}\ \bibnamefont
  {Kovacs}},\ }\href@noop {} {\bibfield  {journal} {\bibinfo  {journal} {Adv.
  Polym. Sci.}\ }\textbf {\bibinfo {volume} {3}},\ \bibinfo {pages} {394}
  (\bibinfo {year} {1963})}\BibitemShut {NoStop}%
\bibitem [{\citenamefont {Kovacs}\ \emph {et~al.}(1979)\citenamefont {Kovacs},
  \citenamefont {Aklonis}, \citenamefont {Hutchinson},\ and\ \citenamefont
  {Ramos}}]{KAHR79}%
  \BibitemOpen
  \bibfield  {author} {\bibinfo {author} {\bibfnamefont {A.~J.}\ \bibnamefont
  {Kovacs}}, \bibinfo {author} {\bibfnamefont {J.~J.}\ \bibnamefont {Aklonis}},
  \bibinfo {author} {\bibfnamefont {J.~M.}\ \bibnamefont {Hutchinson}}, \ and\
  \bibinfo {author} {\bibfnamefont {A.~R.}\ \bibnamefont {Ramos}},\ }\href@noop
  {} {\bibfield  {journal} {\bibinfo  {journal} {J. Polym. Sci.}\ }\textbf
  {\bibinfo {volume} {17}},\ \bibinfo {pages} {1097} (\bibinfo {year}
  {1979})}\BibitemShut {NoStop}%
\bibitem [{\citenamefont {Tu}\ \emph {et~al.}(1998)\citenamefont {Tu},
  \citenamefont {Toner},\ and\ \citenamefont {Ulm}}]{TTU98}%
  \BibitemOpen
  \bibfield  {author} {\bibinfo {author} {\bibfnamefont {Y.}~\bibnamefont
  {Tu}}, \bibinfo {author} {\bibfnamefont {J.}~\bibnamefont {Toner}}, \ and\
  \bibinfo {author} {\bibfnamefont {M.}~\bibnamefont {Ulm}},\ }\href {\doibase
  10.1103/PhysRevLett.80.4819} {\bibfield  {journal} {\bibinfo  {journal}
  {Phys. Rev. Lett.}\ }\textbf {\bibinfo {volume} {80}},\ \bibinfo {pages}
  {4819} (\bibinfo {year} {1998})}\BibitemShut {NoStop}%
\bibitem [{\citenamefont {Toner}\ \emph {et~al.}(2005)\citenamefont {Toner},
  \citenamefont {Tu},\ and\ \citenamefont {Ramaswamy}}]{TTR05}%
  \BibitemOpen
  \bibfield  {author} {\bibinfo {author} {\bibfnamefont {J.}~\bibnamefont
  {Toner}}, \bibinfo {author} {\bibfnamefont {Y.}~\bibnamefont {Tu}}, \ and\
  \bibinfo {author} {\bibfnamefont {S.}~\bibnamefont {Ramaswamy}},\ }\href
  {\doibase https://doi.org/10.1016/j.aop.2005.04.011} {\bibfield  {journal}
  {\bibinfo  {journal} {Ann. Phys.}\ }\textbf {\bibinfo {volume} {318}},\
  \bibinfo {pages} {170 } (\bibinfo {year} {2005})},\ \bibinfo {note} {special
  Issue}\BibitemShut {NoStop}%
\bibitem [{\citenamefont {Toner}(2012)}]{Toner12}%
  \BibitemOpen
  \bibfield  {author} {\bibinfo {author} {\bibfnamefont {J.}~\bibnamefont
  {Toner}},\ }\href {\doibase 10.1103/PhysRevLett.108.088102} {\bibfield
  {journal} {\bibinfo  {journal} {Phys. Rev. Lett.}\ }\textbf {\bibinfo
  {volume} {108}},\ \bibinfo {pages} {088102} (\bibinfo {year}
  {2012})}\BibitemShut {NoStop}%
\bibitem [{\citenamefont {Berthier}\ and\ \citenamefont
  {Bouchaud}(2002)}]{BB02}%
  \BibitemOpen
  \bibfield  {author} {\bibinfo {author} {\bibfnamefont {L.}~\bibnamefont
  {Berthier}}\ and\ \bibinfo {author} {\bibfnamefont {J.-P.}\ \bibnamefont
  {Bouchaud}},\ }\href@noop {} {\bibfield  {journal} {\bibinfo  {journal}
  {Phys. Rev. B}\ }\textbf {\bibinfo {volume} {66}},\ \bibinfo {pages} {054404}
  (\bibinfo {year} {2002})}\BibitemShut {NoStop}%
\bibitem [{\citenamefont {Cugliandolo}\ \emph {et~al.}(2004)\citenamefont
  {Cugliandolo}, \citenamefont {Lozano},\ and\ \citenamefont {Lozza}}]{CLL04}%
  \BibitemOpen
  \bibfield  {author} {\bibinfo {author} {\bibfnamefont {L.}~\bibnamefont
  {Cugliandolo}}, \bibinfo {author} {\bibfnamefont {G.}~\bibnamefont {Lozano}},
  \ and\ \bibinfo {author} {\bibfnamefont {H.}~\bibnamefont {Lozza}},\
  }\href@noop {} {\bibfield  {journal} {\bibinfo  {journal} {Eur. Phys. J. B}\
  }\textbf {\bibinfo {volume} {41}},\ \bibinfo {pages} {87} (\bibinfo {year}
  {2004})}\BibitemShut {NoStop}%
\bibitem [{\citenamefont {Arenzon}\ and\ \citenamefont
  {Sellitto}(2004)}]{AS04}%
  \BibitemOpen
  \bibfield  {author} {\bibinfo {author} {\bibfnamefont {J.~J.}\ \bibnamefont
  {Arenzon}}\ and\ \bibinfo {author} {\bibfnamefont {M.}~\bibnamefont
  {Sellitto}},\ }\href@noop {} {\bibfield  {journal} {\bibinfo  {journal} {Eur.
  Phys. J. B}\ }\textbf {\bibinfo {volume} {42}},\ \bibinfo {pages} {543}
  (\bibinfo {year} {2004})}\BibitemShut {NoStop}%
\bibitem [{\citenamefont {Ruiz-Garc{\'\i}a}\ and\ \citenamefont
  {Prados}(2014)}]{RP14}%
  \BibitemOpen
  \bibfield  {author} {\bibinfo {author} {\bibfnamefont {M.}~\bibnamefont
  {Ruiz-Garc{\'\i}a}}\ and\ \bibinfo {author} {\bibfnamefont {A.}~\bibnamefont
  {Prados}},\ }\href@noop {} {\bibfield  {journal} {\bibinfo  {journal} {Phys.
  Rev. E}\ }\textbf {\bibinfo {volume} {89}},\ \bibinfo {pages} {012140}
  (\bibinfo {year} {2014})}\BibitemShut {NoStop}%
\bibitem [{\citenamefont {Mossa}\ and\ \citenamefont {Sciortino}(2004)}]{MS04}%
  \BibitemOpen
  \bibfield  {author} {\bibinfo {author} {\bibfnamefont {S.}~\bibnamefont
  {Mossa}}\ and\ \bibinfo {author} {\bibfnamefont {F.}~\bibnamefont
  {Sciortino}},\ }\href@noop {} {\bibfield  {journal} {\bibinfo  {journal}
  {Phys. Rev. Lett.}\ }\textbf {\bibinfo {volume} {92}},\ \bibinfo {pages}
  {045504} (\bibinfo {year} {2004})}\BibitemShut {NoStop}%
\bibitem [{\citenamefont {Prados}\ and\ \citenamefont {Trizac}(2014)}]{PT14}%
  \BibitemOpen
  \bibfield  {author} {\bibinfo {author} {\bibfnamefont {A.}~\bibnamefont
  {Prados}}\ and\ \bibinfo {author} {\bibfnamefont {E.}~\bibnamefont
  {Trizac}},\ }\href@noop {} {\bibfield  {journal} {\bibinfo  {journal} {Phys.
  Rev. Lett.}\ }\textbf {\bibinfo {volume} {112}},\ \bibinfo {pages} {198001}
  (\bibinfo {year} {2014})}\BibitemShut {NoStop}%
\bibitem [{\citenamefont {Brey}\ \emph {et~al.}(2014)\citenamefont {Brey},
  \citenamefont {de~Soria}, \citenamefont {Maynar},\ and\ \citenamefont
  {Buz{\'o}n}}]{BGMB14}%
  \BibitemOpen
  \bibfield  {author} {\bibinfo {author} {\bibfnamefont {J.~J.}\ \bibnamefont
  {Brey}}, \bibinfo {author} {\bibfnamefont {M.~G.}\ \bibnamefont {de~Soria}},
  \bibinfo {author} {\bibfnamefont {P.}~\bibnamefont {Maynar}}, \ and\ \bibinfo
  {author} {\bibfnamefont {V.}~\bibnamefont {Buz{\'o}n}},\ }\href@noop {}
  {\bibfield  {journal} {\bibinfo  {journal} {Phys. Rev. E}\ }\textbf {\bibinfo
  {volume} {90}},\ \bibinfo {pages} {032207} (\bibinfo {year}
  {2014})}\BibitemShut {NoStop}%
\bibitem [{\citenamefont {Trizac}\ and\ \citenamefont {Prados}(2014)}]{TP14}%
  \BibitemOpen
  \bibfield  {author} {\bibinfo {author} {\bibfnamefont {E.}~\bibnamefont
  {Trizac}}\ and\ \bibinfo {author} {\bibfnamefont {A.}~\bibnamefont
  {Prados}},\ }\href@noop {} {\bibfield  {journal} {\bibinfo  {journal} {Phys.
  Rev. E}\ }\textbf {\bibinfo {volume} {90}},\ \bibinfo {pages} {012204}
  (\bibinfo {year} {2014})}\BibitemShut {NoStop}%
\bibitem [{\citenamefont {Prados}\ and\ \citenamefont {Brey}(2010)}]{PB10}%
  \BibitemOpen
  \bibfield  {author} {\bibinfo {author} {\bibfnamefont {A.}~\bibnamefont
  {Prados}}\ and\ \bibinfo {author} {\bibfnamefont {J.}~\bibnamefont {Brey}},\
  }\href@noop {} {\bibfield  {journal} {\bibinfo  {journal} {J. Stat. Mech.}\
  }\textbf {\bibinfo {volume} {2010}},\ \bibinfo {pages} {P02009} (\bibinfo
  {year} {2010})}\BibitemShut {NoStop}%
\bibitem [{\citenamefont {Deffuant}\ \emph {et~al.}(2000)\citenamefont
  {Deffuant}, \citenamefont {Neau}, \citenamefont {Amblard},\ and\
  \citenamefont {Weisbuch}}]{DNAW00}%
  \BibitemOpen
  \bibfield  {author} {\bibinfo {author} {\bibfnamefont {G.}~\bibnamefont
  {Deffuant}}, \bibinfo {author} {\bibfnamefont {D.}~\bibnamefont {Neau}},
  \bibinfo {author} {\bibfnamefont {F.}~\bibnamefont {Amblard}}, \ and\
  \bibinfo {author} {\bibfnamefont {G.}~\bibnamefont {Weisbuch}},\ }\href@noop
  {} {\bibfield  {journal} {\bibinfo  {journal} {Adv. Compl. Syst.}\ }\textbf
  {\bibinfo {volume} {3}},\ \bibinfo {pages} {87} (\bibinfo {year}
  {2000})}\BibitemShut {NoStop}%
\bibitem [{\citenamefont {Hegselmann}\ and\ \citenamefont
  {Krause}(2002)}]{HK02}%
  \BibitemOpen
  \bibfield  {author} {\bibinfo {author} {\bibfnamefont {R.}~\bibnamefont
  {Hegselmann}}\ and\ \bibinfo {author} {\bibfnamefont {U.}~\bibnamefont
  {Krause}},\ }\href@noop {} {\bibfield  {journal} {\bibinfo  {journal}
  {JASSS}\ }\textbf {\bibinfo {volume} {5}} (\bibinfo {year}
  {2002})}\BibitemShut {NoStop}%
\bibitem [{\citenamefont {Lu}\ \emph {et~al.}(2013)\citenamefont {Lu},
  \citenamefont {Bi}, \citenamefont {Liu}, \citenamefont {Wu}, \citenamefont
  {Xing},\ and\ \citenamefont {Yeow}}]{lu_13}%
  \BibitemOpen
  \bibfield  {author} {\bibinfo {author} {\bibfnamefont {S.}~\bibnamefont
  {Lu}}, \bibinfo {author} {\bibfnamefont {W.}~\bibnamefont {Bi}}, \bibinfo
  {author} {\bibfnamefont {F.}~\bibnamefont {Liu}}, \bibinfo {author}
  {\bibfnamefont {X.}~\bibnamefont {Wu}}, \bibinfo {author} {\bibfnamefont
  {B.}~\bibnamefont {Xing}}, \ and\ \bibinfo {author} {\bibfnamefont {E.~K.}\
  \bibnamefont {Yeow}},\ }\href@noop {} {\bibfield  {journal} {\bibinfo
  {journal} {Phys. Rev. Lett.}\ }\textbf {\bibinfo {volume} {111}},\ \bibinfo
  {pages} {208101} (\bibinfo {year} {2013})}\BibitemShut {NoStop}%
\bibitem [{Note1()}]{Note1}%
  \BibitemOpen
  \bibinfo {note} {\label {fn:supp}See supplemental material at pages
  6ff.}\BibitemShut {Stop}%
\bibitem [{Note2()}]{Note2}%
  \BibitemOpen
  \bibinfo {note} {\label {foot:1}Since the density $M=0.2$ is not that small
  the description by the kinetic theory \protect \textup {\hbox {\mathsurround
  \z@ \protect \normalfont (\ignorespaces \ref {eq:timeevolutionfourier}\unskip
  \@@italiccorr )}} is not perfect and the steady state of the real system is
  slightly different.\label {1} To compare kinetic theory and simulations we
  used slightly different noise strengths for the simulation, such that the
  steady state value of the order parameter coincides with the kinetic theory
  for all three noise strengths. Simulation parameters are $\eta _1=0.38$,
  $\eta _2=0.46$, $\eta _f=0.395$.}\BibitemShut {Stop}%
\bibitem [{Note3()}]{Note3}%
  \BibitemOpen
  \bibinfo {note} {Analogously to \cite {Note2} simulation noise strengths
  differ minimally: $\eta _1=0.34$, $\eta _2=0.197$, $\eta
  _f=0.29$.}\BibitemShut {Stop}%
\bibitem [{\citenamefont {Bouchaud}\ \emph {et~al.}(2001)\citenamefont
  {Bouchaud}, \citenamefont {Dupuis}, \citenamefont {Hammann},\ and\
  \citenamefont {Vincent}}]{BDHV01}%
  \BibitemOpen
  \bibfield  {author} {\bibinfo {author} {\bibfnamefont {J.-P.}\ \bibnamefont
  {Bouchaud}}, \bibinfo {author} {\bibfnamefont {V.}~\bibnamefont {Dupuis}},
  \bibinfo {author} {\bibfnamefont {J.}~\bibnamefont {Hammann}}, \ and\
  \bibinfo {author} {\bibfnamefont {E.}~\bibnamefont {Vincent}},\ }\href@noop
  {} {\bibfield  {journal} {\bibinfo  {journal} {Phys. Rev. B}\ }\textbf
  {\bibinfo {volume} {65}},\ \bibinfo {pages} {024439} (\bibinfo {year}
  {2001})}\BibitemShut {NoStop}%
\bibitem [{Note4()}]{Note4}%
  \BibitemOpen
  \bibinfo {note} {To determine the largest eigenvalue of $[H_{\eta }^{t}]^{L}$
  we fix a small parameter $\delta $, here $\delta =0.01$. We start with the
  initial vector $\protect \mathbf {x}$ given by $x_1=1$ and $x_i=0$ for $i>1$.
  We then replace $\protect \mathbf {x}$ by $H_{\eta }^{t}(\delta \cdot
  \protect \mathbf {x})/ || H_{\eta }^{t}(\delta \cdot \protect \mathbf {x})
  ||$, such that always $|| \protect \mathbf {x} ||=1$. This step is repeated
  many times. In this way all eigenvectors but the one that corresponds to the
  largest eigenvalue decay. Eventually $\protect \mathbf {x}$ converges and the
  largest eigenvalue is obtained by $\lambda _{1}^{\eta , t}=||H_{\eta
  }^{t}(\delta \cdot \protect \mathbf {x}) ||/\delta $ and $\protect \mathbf
  {x}$ is the corresponding eigenvector.}\BibitemShut {Stop}%
\end{thebibliography}%

\begin{widetext}

\makeatletter

\setcounter{equation}{0}
\setcounter{figure}{0}
\renewcommand{\theequation}{S.\arabic{equation}}
\renewcommand{\thefigure}{S\arabic{figure}}

\makeatother

\newpage

\title{Supplemental Material to 'Giant Kovacs-Like Memory Effect for Active Particles'}

\begin{center}
	\large{\textbf{Supplemental Material to 'Giant Kovacs-Like Memory Effect for Active Particles'}}
\end{center}

\section{Nonlinear Theory\label{sec:derivation}}

In the following we give the derivation of Eq.~(\ref{eq:kovacshump}).
Consider a family of maps $F_{\eta}: \mathbb{R}^{n} \rightarrow \mathbb{R}^{n}$. 
In the present system the map $F$ is given by Eq.~(\ref{eq:timeevolutionfourier}).
We assume that each map has a single globally attractive stable fixed point $\mathbf{x}^{*}_{\eta}$ such that $F_{\eta}(\mathbf{x}^{*}_{\eta})=\mathbf{x}^{*}_{\eta}$.
We define the shifted coordinate $\mathbf{y}$ and map $G_{\eta}$ by
\begin{align}
	\mathbf{y}&:= \mathbf{x}-\mathbf{x}^{*}_{\eta_f},
	\qquad
	G_{\eta}(\mathbf{y}):= F_{\eta}(\mathbf{y} + \mathbf{x}^{*}_{\eta_f}) - \mathbf{x}^{*}_{\eta_f}
	\label{eq:shifted}
\end{align}
such that
\begin{align}
	G_{\eta_f}(0)=0.
	\label{eq:fpshiftedetastar}
\end{align}
We denote the fixed point of $G_{\eta}$ by
\begin{align}
	\mathbf{y}^{*}_{\eta}:= \mathbf{x}^{*}_{\eta}- \mathbf{x}^{*}_{\eta_f}.
	\label{eq:fpshifted}
\end{align}
Furthermore we define the relaxation map
\begin{align}
	H_{\eta}(\mathbf{y}-\mathbf{y}^{*}_{\eta}):= G_{\eta}(\mathbf{y})-\mathbf{y}_{\eta_f}
	\label{eq:relaxmap}
\end{align}
which has the property that $\lim_{t\rightarrow \infty}H^{t}_{\eta}(\mathbf{y})=0$.
Let
\begin{align}
	\mathbf{y}(t=0)=\mathbf{y}^{*}_{\eta_1},
	\label{eq:initialconditiony}
\end{align}
for $0<t\le t_{w}$
\begin{align}
	\mathbf{y}(t)&= G^{t}_{\eta_2}(\mathbf{y}(t=0))=G^{t}_{\eta_{2}}(\mathbf{y}^{*}_{\eta_{1}})
	= \mathbf{y}^{*}_{\eta_2}+H^{t}_{\eta_2}(\mathbf{y}^{*}_{\eta_1}-\mathbf{y}^{*}_{\eta_2})
	\label{eq:timeevolution1}
\end{align}
and for $t>t_{w}$
\begin{align}
	\mathbf{y}(t)&=G_{\eta^{*}}^{t-t_{w}}(\mathbf{y}(t_{w}))=G^{t-t_{w}}_{\eta_f}(\mathbf{y}^{*}_{\eta_{2}}+H^{t_{w}}_{\eta_2}(\mathbf{y}^{*}_{\eta_1}-\mathbf{y}^{*}_{\eta_2}))
	=H^{t-t_{w}}_{\eta_f}(\mathbf{y}^{*}_{\eta_{2}}+H^{t_{w}}_{\eta_2}(\mathbf{y}^{*}_{\eta_1}-\mathbf{y}^{*}_{\eta_2}))
	\label{eq:timeevolution2}
\end{align}
where the last line follows from $\mathbf{y}^{*}_{\eta_f}=0$.
So far we have only formulated the dynamics of the system and Eq.~(\ref{eq:timeevolution2}) is exact.
Assuming that the changes of the noise strength $\eta_1-\eta_f$ and $\eta_2-\eta_f$ are small, of order $\varepsilon$, and neglecting all terms of higher order than $\varepsilon$ we rederive Eq.~(\ref{eq:lintheory}), cf. Sec.~\ref{sec:lintheory}.

In a different approach we keep the full dependence on the changes of the noise strength by now.
We replace $H_{\eta_f}^{t-t_{w}}$ by its linearization $\big[ H_{\eta_f}^{t-t_{w}}\big]^{L}$ such that Eq.~(\ref{eq:timeevolution2}) becomes
\begin{align}
	\mathbf{y}(t)&\approx \big[ H^{t-t_{w}}_{\eta_f} \big]^{L}(\mathbf{y}^{*}_{\eta_2}) + \big[ H^{t-t_{w}}_{\eta_f} \big]^{L} (H^{t_{w}}_{\eta_2}(\mathbf{y}^{*}_{\eta_1}-\mathbf{y}^{*}_{\eta_{2}}))
	\approx H^{t-t_{w}}_{\eta_f} (\mathbf{y}^{*}_{\eta_2}) + \big[ H^{t-t_{w}}_{\eta_f} \big]^{L} (H^{t_{w}}_{\eta_2}(\mathbf{y}^{*}_{\eta_1}-\mathbf{y}^{*}_{\eta_{2}}))
	\notag
	\\
	&= \mathbf{y}_{2f}(t-t_{w}) + \big[ H^{t-t_{w}}_{\eta_f} \big]^{L} (H^{t_{w}}_{\eta_2}(\mathbf{y}^{*}_{\eta_1}-\mathbf{y}^{*}_{\eta_{2}}))
	\label{eq:timeevolution2lin}
\end{align}
Furthermore we also replace the map $H^{t_{w}}_{\eta_2}$ by its linearization $\big[H^{t_{w}}_{\eta_2}\big]^{L}$ and hence Eq.~(\ref{eq:timeevolution2lin}) becomes
\begin{align}
	\mathbf{y}(t)=& \mathbf{y}_{2f}(t-t_{w}) + \big[ H^{t-t_{w}}_{\eta_f} \big]^{L} (\big[H^{t_{w}}_{\eta_2}\big]^L(\mathbf{y}^{*}_{\eta_1})) 
	- \big[ H^{t-t_{w}}_{\eta_f} \big]^{L} (\big[H^{t_{w}}_{\eta_2}\big]^L(\mathbf{y}^{*}_{\eta_2})).
	\label{eq:timeevolution2linlin}
\end{align}
We denote the eigenvalues of $\big[H^{t}_{\eta}\big]^L$ by $\lambda_{1}^{\eta, t} \ge \lambda_{2}^{\eta, t} \ge \dots \ge \lambda_{n}^{\eta, t}$ and the corresponding eigenvectors by $\mathbf{v}_{1}^{\eta, t}, \mathbf{v}_{2}^{\eta, t}, \dots, \mathbf{v}_{n}^{\eta, t}$. 
For $t=t_{w}$ we assume a separation of time scales, property (i) which is checked numerically below, that means 
\begin{align}
\lambda_{1}^{\eta, t_w} \gg \lambda_{2}^{\eta, t_w}.
\label{eq:eigenvectorseparation}
\end{align}
Therefore we can assume that at $t_w$ only the first eigenvector is relevant and all others are much smaller, thus
\begin{align}
	\big[H^{t_w}_{\eta_2}\big]^L(\mathbf{y}) &= \sum_{k=1}^{n} \lambda_{k}^{\eta_2, t_w}  (\mathbf{y}\cdot \mathbf{v}_{k}^{\eta_2, t_w}) \mathbf{v}_{k}^{\eta_2, t_w}
	\approx  \lambda_{1}^{\eta_2, t_w}  (\mathbf{y}\cdot \mathbf{v}_{1}^{\eta_2, t_w}) \mathbf{v}_{1}^{\eta_2, t_w}
	\label{eq:onlyfirsteigenvector}
\end{align}
We further assume that the eigenvector belonging to the largest eigenvalue is not sensitive to moderate changes in $\eta$ and $t$, property (ii) which is checked numerically below. That means we assume that
\begin{align}
	\mathbf{v}_{1}^{\eta_2, t_w} \approx \mathbf{v}_{1}^{\eta_f, \tilde{t}}
	\label{eq:equaleigenvectors}
\end{align}
when $\tilde{t}$ has the same order of magnitude as $t_w$. Then we choose $\tilde{t}$ such that
\begin{align}
	\lambda_{1}^{\eta_2, t_w}=\lambda_{1}^{\eta_f, \tilde{t}}
	\label{eq:eigenvaluecondition}
\end{align}
and hence from Eq.~(\ref{eq:onlyfirsteigenvector}) we obtain
\begin{align}
	\big[H^{t_w}_{\eta_2}\big]^L(\mathbf{y}) 
	&\approx  \lambda_{1}^{\eta_2, t_w}  (\mathbf{y}\cdot \mathbf{v}_{1}^{\eta_2, t_w}) \mathbf{v}_{1}^{\eta_2, t_w}
	\approx  \lambda_{1}^{\eta_f, \tilde{t}}  (\mathbf{y}\cdot \mathbf{v}_{1}^{\eta_f, \tilde{t}}) \mathbf{v}_{1}^{\eta_f, \tilde{t}}
	\approx \big[H^{\tilde{t}}_{\eta_f}\big]^L(\mathbf{y}).
	\label{eq:timeshift}
\end{align}
Plugging this expression into Eq.~(\ref{eq:timeevolution2linlin}) we obtain
\begin{align}
	\mathbf{y}(t)=& \mathbf{y}_{2f}(t-t_{w}) + \big[ H^{t-t_{w}}_{\eta_f} \big]^{L} (\big[H^{\tilde{t}}_{\eta_f}\big]^L(\mathbf{y}^{*}_{\eta_1})) 
	- \big[ H^{t-t_{w}}_{\eta_f} \big]^{L} (\big[H^{\tilde{t}}_{\eta_f}\big]^L(\mathbf{y}^{*}_{\eta_2}))
	\notag
	\\
	\approx & \mathbf{y}_{2f}(t-t_{w}) + H^{t-t_{w}}_{\eta_f} (H^{\tilde{t}}_{\eta_f}(\mathbf{y}^{*}_{\eta_1})) 
	-  H^{t-t_{w}}_{\eta_f} (\big[H^{\tilde{t}}_{\eta_f}(\mathbf{y}^{*}_{\eta_2}))
	\notag
	\\
	=& \mathbf{y}_{2f}(t-t_{w}) + \mathbf{y}_{1f}(t+\tilde{t}-t_{w}) 
	- \mathbf{y}_{2f}(t+\tilde{t}-t_{w})
	\label{eq:timeevolution2linlinapprox}
\end{align}
Resubstituting $\mathbf{y}$ according to Eq.~(\ref{eq:shifted}) and introducing the abbreviation
\begin{align}
	\hat{t}:=t_w-\tilde{t}
	\label{eq:that}
\end{align}
we obtain Eq.~(\ref{eq:kovacshump}).

The basic assumptions in the derivation of Eq.~(\ref{eq:kovacshump}) are properties i) and ii) that are mathematically given by the conditions (\ref{eq:eigenvectorseparation}) and (\ref{eq:equaleigenvectors}).
In the letter, we have argued intuitively that these assumptions should be valid in the present system.
For the example presented in Fig.~\ref{fig:giant} we can also explicitly verify those conditions numerically.
The time evolution map $F$ is explicitly given by Eq.~(\ref{eq:timeevolutionfourier}).
Evolving the system for a very long time we obtain the stable fixed point of the map.
Thus we can also evolve the map $H$, cf. Eq.~(\ref{eq:relaxmap}).

To verify condition (\ref{eq:eigenvectorseparation}) we need to obtain the eigenvalues $\lambda_{i}^{\eta, t}$ of $[H_{\eta}^{t}]^{L}$.
As a rough estimate we can instead calculate the eigenvalues $\lambda_{i}^{\eta}$ of $[H_{\eta}]^{L}$ and assume $\lambda_{i}^{\eta, t}\sim (\lambda_{i}^{\eta})^{t}$.
Given the fixed point of the system, the linearization of $H_{\eta}$ can be performed analytically.
For the parameters of Fig.~\ref{fig:giant} we obtain for the two largest eigenvalues
\begin{align}
	(\lambda_{1}^{\eta_2})^{t_w} \approx 8.2\times 10^{-2} \gg (\lambda_{2}^{\eta_2})^{t_w} \approx 1.6 \times 10^{-7}.
	\label{eq:twolargesteigenvalues}
\end{align}
Thus our assumption (\ref{eq:eigenvectorseparation}) is valid.

To verify the condition (\ref{eq:equaleigenvectors}) we use a robust numerical procedure to obtain the largest eigenvalue and the corresponding eigenvector of $[H_{\eta}^{t}]^{L}$ \footnote{
To determine the largest eigenvalue of $[H_{\eta}^{t}]^{L}$ we fix a small parameter $\delta$, here $\delta=0.01$.
We start with the initial vector $\mathbf{x}$ given by $x_1=1$ and $x_i=0$ for $i>1$.
We then replace $\mathbf{x}$ by $H_{\eta}^{t}(\delta\cdot\mathbf{x})/  || H_{\eta}^{t}(\delta\cdot\mathbf{x}) ||$, such that always $|| \mathbf{x} ||=1$.
This step is repeated many times. In this way all eigenvectors but the one that corresponds to the largest eigenvalue decay.
Eventually $\mathbf{x}$ converges and the largest eigenvalue is obtained by $\lambda_{1}^{\eta, t}=||H_{\eta}^{t}(\delta\cdot\mathbf{x}) ||/\delta$ and $\mathbf{x}$ is the corresponding eigenvector.
}.
For parameters like in Fig.~\ref{fig:giant} we find that Eq.~(\ref{eq:eigenvaluecondition}) is valid for $\tilde{t}=408$.
Knowing $\tilde{t}$ we can calculate the scalar product of the normalized eigenvectors $\mathbf{v}_{1}^{\eta_2, t_w} \cdot \mathbf{v}_{1}^{\eta_f, \tilde{t}} \approx 0.963$ which indicates that the eigenvector $\mathbf{v}_1$ did not change a lot and Eq.~\eqref{eq:equaleigenvectors} is a reasonable assumption.

As we have predicted the value of $\tilde{t}$ we can also calculate the time shift $\hat{t}=337$ according to Eq.~(\ref{eq:that}).
In Fig.~\ref{fig:giant} we used the time shift $\hat{t}=332$ that was chosen such that the Kovacs-hump has the correct value at $t=t_w$.
Hence our numerical prediction for $\hat{t}$ deviates by about $1\%$ from the correct value.
This is not surprising since the assumption \eqref{eq:equaleigenvectors} is not satisfied perfectly and also the linearization of $H_{\eta}^{t}$ might introduce some deviations.

Our result \eqref{eq:kovacshump} describes a relation between vectors of all Fourier modes. Thus it is holding not only for the polar order parameter $\Psi= \pi g_1$ which corresponds to the first Fourier mode but it is much more general.
In Fig.~\ref{fig:secondmode} we show the relaxation curves and the Kovacs-hump of the second Fourier mode following the same Kovacs-protocol as the order parameter in Fig.~\ref{fig:giant}.
We find that the relation \eqref{eq:kovacshump} fits the data well using the same time shift $\hat{t}$ as in Fig.~\ref{fig:giant}.
\begin{figure}
	\includegraphics{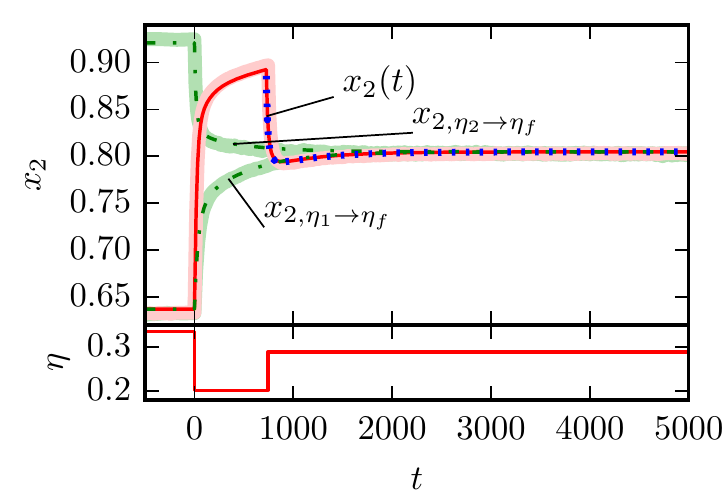}
	\caption{
	The relaxation curves $x_{2, \eta_1 \rightarrow \eta_f}(t)$, $x_{2, \eta_2 \rightarrow \eta_f}(t)$ (dash-dotted line) and 
the Kovacs-hump $x_2(t)$ (solid red line) of the second Fourier mode $g_2$ obtained from the kinetic theory \eqref{eq:timeevolutionfourier} are in good agreement with agent-based simulations (light green and light red thick line, respectively). 
	The nonlinear theory \eqref{eq:kovacshump} with the same time shift as in Fig.~\ref{fig:giant}, $\hat{t}=332$, (dotted blue line) fits well.
	Parameters are as in Fig.~\ref{fig:giant}.
	\label{fig:secondmode}}
\end{figure}

\section{Linear Theory\label{sec:lintheory}}
We assume that 
\begin{align}
	\eta_1 - \eta_f&= a_1 \varepsilon,
	\qquad
	\eta_2 - \eta_f= a_2 \varepsilon,
	\label{eq:a2smallepsilon}
\end{align}
where $\varepsilon$ is small and $a_1$ and $a_2$ are of similar size, $a_{1}, a_{2} \sim 1$.
We expand the fixed points
\begin{align}
	\mathbf{y}^{*}_{\eta_1} &= \varepsilon a_1 \mathbf{y}_{0}^{*} + \mathcal{O}(\varepsilon^2),
	\qquad
	\mathbf{y}^{*}_{\eta_2} = \varepsilon a_2 \mathbf{y}_{0}^{*} + \mathcal{O}(\varepsilon^2)
	\label{eq:fixedpointsmallepsilon2}
\end{align}
for some constant vector $\mathbf{y}^{*}_{0}$.
Inserting these fixed points into the exact expression \eqref{eq:timeevolution2} we obtain
\begin{align}
	\mathbf{y}(t) &= H^{t-t_w}_{\eta_f}( \varepsilon a_2 \mathbf{y}_{0}^{*} + H_{\eta_2}^{t_w}(\varepsilon(a_1-a_2) \mathbf{y}_{0}^{*} + \mathcal{O}(\varepsilon^2)) + \mathcal{O}(\varepsilon^2) )
	\notag
	\\
	 &= H^{t-t_w}_{\eta_f}( \varepsilon [ a_2 \mathbf{y}_{0}^{*} + (a_1-a_2) ([H_{\eta_2}]^L)^{t_w}(\mathbf{y}_{0}^{*} )] + \mathcal{O}(\varepsilon^2) ),
	\label{eq:linearization1}
\end{align}
where $[H_{\eta_2}]^L$ denotes the linearization of $H_{\eta_2}$.
The linear map $[H_{\eta_2}]^{L}$ depends on $\varepsilon$ via $\eta_2$ and thus we can expand it for small $\varepsilon$.
Only the leading term is necessary since we neglect terms of order $\varepsilon^2$. Thus we obtain
\begin{align}
	\mathbf{y}(t) 
	&= H^{t-t_w}_{\eta_f}( \varepsilon [ a_2 \mathbf{y}_{0}^{*} + (a_1-a_2) ([H_{\eta_f}]^L)^{t_w}(\mathbf{y}_{0}^{*} ) ] + \mathcal{O}(\varepsilon^2) ).
	\label{eq:linearization2}
\end{align}
Since the map $[H_{\eta_2}]^{L}$ is applied $t_w$ times, terms of order $\varepsilon^2$ come with a prefactor of $t_w$. 
Therefore they can be neglected only if $1/\varepsilon \gg t_w$.
For not too small changes in $\eta$ and long waiting times this assumption is not maintainable and the linear theory cannot be applied.
This is the reason why Eq.~(\ref{eq:lintheory}) fails completely for long waiting times.

Continuing the expansion of Eq.~\eqref{eq:linearization2} for small $\varepsilon$ we obtain
\begin{align}
	\mathbf{y}(t) 
	&= \varepsilon ([H_{\eta_f}]^L)^{t-t_w}( a_2 \mathbf{y}_{0}^{*} + (a_1-a_2) ([H_{\eta_f}]^L)^{t_w}(\mathbf{y}_{0}^{*} )  )+ \mathcal{O}(\varepsilon^2)
	\notag
	\\
	&= \frac{a_2}{a_1} ([H_{\eta_f}]^L)^{t-t_w}( \varepsilon a_1 \mathbf{y}_{0}^{*}   )+ \frac{a_1-a_2}{a_1} ([H_{\eta_f}]^L)^{t_w}( \varepsilon a_1 \mathbf{y}_{0}^{*}   ) + \mathcal{O}(\varepsilon^2)
	\notag
	\\
	&= \frac{a_2}{a_1} \mathbf{y}_{1f}(t-t_w) + \frac{a_1-a_2}{a_1} \mathbf{y}_{1f}(t) + \mathcal{O}(\varepsilon^2)
	\label{eq:linearization3}
\end{align}
Evaluating this equation at $t=0$, resubstituting $\mathbf{x}$ according to Eq.~(\ref{eq:fpshifted}) and observing only $x_1=\Psi/\pi$ we find $\gamma$ according to Eq.~(\ref{eq:gamma}) as
\begin{align}
	\gamma = \frac{a_2}{a_2-a_1}.
	\label{eq:gamma2}
\end{align}
With this abbreviation Eq.~\eqref{eq:linearization3} becomes
\begin{align}
	\mathbf{y}(t)&= \frac{1}{1-\gamma}\mathbf{y}_{\eta_1\rightarrow \eta_f}(t) - \frac{\gamma}{1-\gamma} \mathbf{y}_{\eta_1 \rightarrow \eta_f}(t-t_{w}).
	\label{eq:linearization4}
\end{align}
From the first vector component of this equation we obtain Eq.~(\ref{eq:lintheory}).
Thus we rederived the linearized theory [43] in the context of time-discrete dynamical systems.

\section{Slow relaxation mechanism \label{sec:pdf}}
In this section, we discuss the slow relaxation mechanism that leads to strong memory effects.
In Fig.~\ref{fig:pdf}, we display snap shots of the angular probability density at times $t=0, 100, 700 < t_w$ following the relaxation from $\eta_2$ to $\eta_f$ which is the first part of the Kovacs-protocol presented in Fig.~\ref{fig:giant}.
\begin{figure}
	\includegraphics{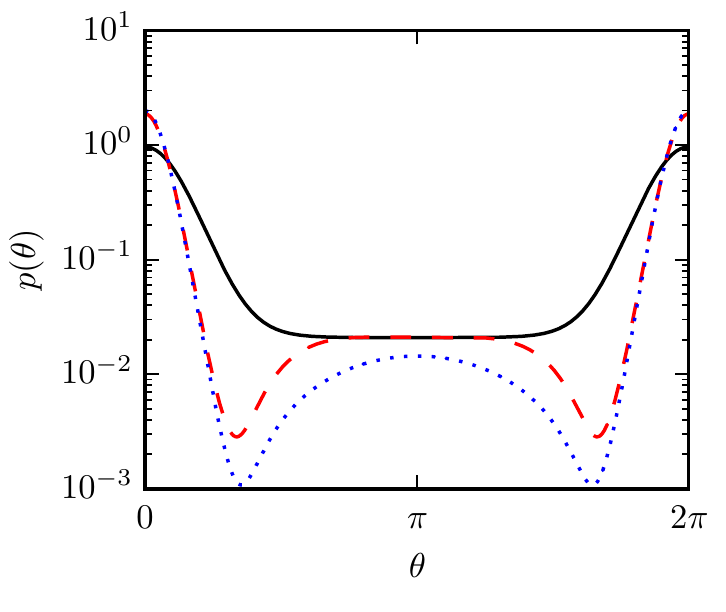}
	\caption{Angular probability density $p(\theta)$ during the Kovacs-protocol at different times before $t_w=742$: $t=0$ (black solid line), $t=100$ (red dashed line) and $t=700$ (blue dotted line). Parameters like in Fig.~\ref{fig:giant}.\label{fig:pdf}}
\end{figure}
The direction of the majority of the particles is centered around $(0 \mod 2\pi)$. 
After $t=100$ this dominating peak is already relaxed towards its steady state shape.
However, there is a small population of particles that move in the opposite direction $\theta=\pi$.
Due to the bounded confidence interaction rule, these particles do not interact with the large group of particles that move in direction $\theta=0$.
Therefore they reorient mainly due to noise.
If the noise strength is small, this process is very slow and it takes a long time until the population moving into direction $\theta=\pi$ has reorientated and the system has reached its steady state.
Apparently this noise driven mechanism is the slowest relaxation mode of the system. 
Moderate changes of the noise strength change the relaxation speed significantly, however, the noise driven reorientation of a minor population remains the slowest relaxation mode of the system.

\section{Coefficients\label{sec:coeff}}
We give the coefficients of Eq.~\eqref{eq:timeevolutionfourier}, cf. Refs. [27, 29] for a derivation.
\begin{align}
	\lambda_{k}(\eta, M) &= \frac{\sin(k\eta/2)}{1+M} \frac{4}{k\eta},
	\\
	A_{k}(M, \alpha) &= \frac{1}{2} + x_{0} M \big\{ \pi-\alpha +\frac{3}{k}\sin(k\alpha/2) 
	- \frac{1}{2k}\sin(k\alpha)  \big\},
	\\
	B_{k}(M, \alpha) &= 
	\frac{M}{2}\alpha \big\{ \sinc[  (k/2-q)\alpha/\pi   ] - \sinc(q\alpha/\pi  ) \big\},
	\\
	C_{k}(M, \alpha) &=  \frac{M}{2}\alpha \big\{ \sinc[ (k/2+q)\alpha/\pi   ] - \sinc(q\alpha/\pi)   \big\}.
\end{align}

\end{widetext}

\end{document}